# Combining Exploratory Analysis and Automated Analysis for Anomaly Detection in Real-Time Data Streams

Ahmed Shah, Ibrahim Abualhaol, Mahmoud Gad, and Michael Weiss

> *Besides black art, there is only automation and mechanization.*
>
> Federico Garcia Lorca (1898–1936)
> Poet, playwright, and theatre director

Security analysts can become overwhelmed with monitoring real-time security information that is important to help them defend their network. They also tend to focus on a limited portion of the alerts, and therefore risk missing important events and links between them. At the heart of the problem is the system that analysts use to detect, explore, and respond to cyber-attacks. Developers of security analysis systems face the challenge of developing a system that can present different sources of information at multiple levels of abstraction, while also creating a system that is intuitive to use. In this article, we examine the complementary nature of exploratory analysis and automated analysis by testing the development of a system that monitors real-time Border Gateway Protocol (BGP) traffic for anomalies that might indicate security threats. BGP is an essential component for supporting the infrastructure of the Internet; however, it is also highly vulnerable and can be hijacked by attackers to propagate spam or launch denial-of-service attacks. Some of the attack scenarios on the BGP infrastructure can be quite elaborate, and it is difficult, if not impossible, to fully automate the detection of such attacks. This article makes two contributions: i) it describes a prototype platform for computing indicators and threat alerts in real time and for visualizing the context of an alert, and ii) it discusses the interaction of exploratory analysis (visualization) and automated analysis. This article is relevant to students, security researchers, and developers who are interested in the development or use of real-time security monitoring systems. They will gain insights into the complementary aspects of automated analysis and exploratory analysis through the development of a real-time streaming system.

## Introduction

Security analysts can easily become overwhelmed with information, which can lead them to neglect critical alerts. This problem is exemplified in the 2013 Target data breach, which is one of the largest security breaches in history: it exposed 40 million credit card accounts and 70 million of the retailer's customer profiles (Krebs, 2013). A forensic analysis of the attack (US Senate, 2014) found that the security monitoring systems put in place by Target had detected many of the key intrusion attempts during the attack; however, Target's analysts were simply overwhelmed by the volume of alerts produced by the system and missed the early warning signs that a major attack was underway.

Indeed, analysts are "bombarded with alerts", receiving so many that "they just don't respond to everything" (Finkle & Heavey, 2014). Also, security analysts tend to focus on a limited portion of the alerts and therefore risk missing important events and relationships (Pierazzi et al., 2016). At the heart of the problem is the system that these analysts use to detect, explore, and respond to unanticipated and anticipated cyber-attacks. Generally speaking, developers of a security analysis platform (such as an intrusion detection system [IDS] or a security information and event management [SIEM] system) can face many challenges. Among them, a key challenge is how much data (e.g., raw traffic data) to present to analysts and to what extent the detection of anomalies should be automated by encoding detection rules into the system.

# Combining Exploratory Analysis and Automated Analysis for Anomaly Detection

*Ahmed Shah, Ibrahim Abualhaol, Mahmoud Gad, and Michael Weiss*

The goal of this article is to examine the complementary nature of exploratory analysis and automated analysis for anomaly detection. For this purpose, we constructed a working prototype of a system to monitor real-time Border Gateway Protocol (BGP) traffic for security threats that combines both aspects. However, the application to BGP, as such, is not at the core of the present work: we simply used it to ground our work in a real-world context. The construction of the prototype produced two outcomes: i) categorization of attacks and indicators related to BGP derived from known threat scenarios and selection of indicators used in the prototype, and ii) an operationalization of the indicators and alerts (automation) and their visualization (exploration).

The findings presented in this article are most relevant to developers of systems for security monitoring. They face the challenge of developing intuitive systems for security analysts who are presented with different sources of information at multiple levels of abstraction (Corona et al., 2009). Developers also need to present this information at a human level of understanding that enables analysts to take appropriate and timely action (Corona et al., 2009). When analysts succeed in detecting "weak signals" (Fink et al., 2005) and acting on them early, their ability to manage security risks is greatly enlarged. It allows them to anticipate future attacks, rather than just reacting as they are detected.

This article is organized into four sections. We first review the literature on modes of analysis, the Border Gateway Protocol (BGP), and indicators and detection techniques for BGP attacks. We then describe the creation of a prototype system that combines exploratory analysis and automated analysis. Next, we examine the trade-off between exploratory analysis and automated analysis. We conclude by discussing lessons from the research that can be applied to the development of real-time security monitoring systems.

## Literature Review

*Modes of analysis*

The automated analysis of network traffic works well for relatively stable environments. However, modern networks are growing in complexity and variability due to their dynamic and heterogeneous nature. This environment can create unstable systems in which the rules used by automated analysis become obsolete over time. Independently of our work, Pierazzi and colleagues (2016) found that a hybrid approach of exploratory analysis and automated analysis is necessary for effective anomaly detection.

Visualizing the observed data can help validate the outcomes of automated analysis. A visual representation of the context of an attack enables verification (Is the automated analysis correct?) and validation (Is the automated analysis meaningful?). Visualization techniques allow people to see and comprehend large amounts of complex data (Riad et al., 2011). Visualization can be used for the iterative improvement of automation rules. It also helps with the further exploration of an alert by an analyst to see what aspects of detection can be automated.

*Border Gateway Protocol*

Management of worldwide Internet traffic is administered by tens of thousands of independent routing domain systems called autonomous systems (AS) (Biersack et al., 2012). An AS can be owned by network operators such as Internet Service Providers (ISPs). The Border Gateway Protocol (BGP) is an inter-domain routing protocol used for managing network reachability information between more than one AS (Rekhter et al., 2006). Although BGP can be thought of as the protocol "that makes the internet work" (Pepelnjak, 2007), it is also considered as "the Internet's biggest security hole" (Zetter, 2008). Malicious actors have the potential to influence BGP to deny service, sniff communications, reroute traffic to malicious networks, and create network instabilities (Meinel, 2008). Abnormal routing behaviour can disrupt global or local bound Internet connectivity and stability (Li et al., 2014; Murphy, 2006).

*Indicators and detection techniques*

In a survey of anomaly detection techniques for BGP data, Al-Musawi (2015) identified key indicators that can be used to detect BGP attacks. Among the most common indicators were the "number of BGP updates" and "AS path length". The most common analytical approaches were time series analysis, machine learning, and statistical pattern recognition including support vector machines, hidden Markov models, and naive Bayes models. Biersack and colleagues (2012) surveyed various visual analytics tools for BGP, including node-link diagrams, rank-charge graphs, timelines, matrices, maps, and charts.

## Creating a Platform that Combines Exploration and Automation

In this section, we describe the outcomes obtained from constructing a prototype of the analysis platform: i) the categorization of attacks and indicators related to BGP, as derived from known threat scenarios and the selection of indicators used in the prototype, and ii) the operationalization of indicators and alerts (automation) and their visualization (exploration). The platform was created strictly using open source technologies such as Apache Spark for real-time stream processing, D3.js and Crossfilter.js for visualization, MongoDB for data storage, Kafka for internal message communication, Flask for creating an external API, and libBGPstream for BGP data stream extraction.

*Categorizing attacks and indicators*
To conduct any kind of security analytics, we need to identify the known types of attacks and their key indicators. One proven way to compile this information is to examine attack cases and extract common attack characteristics and indicators. For this work, we surveyed BGP attack cases that were described in published studies, some of which also included unintentional attacks (e.g., a misconfiguration), as exemplified in Box 1. Preference was given to cases that included a detailed forensic analysis that examined indicators could be used for anomaly detection. Cases were also given priority for in-depth study if the attack dataset was publicly accessible. Out of the 15 cases that were collected, five general scenarios were identified where BGP was used for attacks: distribution of spam, influence of worms, traffic redirection for theft, eavesdropping, and denial-of-service attacks.

**Box 1.** Example scenario of a BGP IP prefix hijack

One widely cited BGP disruption scenario is the IP prefix hijack of YouTube in 2008. This hijack resulted from a foreign telecommunications company misconfiguring their systems: Pakistan Telecom inadvertently prevented users from around the world from accessing YouTube for roughly two hours. Pakistan Telecom was attempting to restrict its users from accessing YouTube. However, they accidentally sent new routing information via BGP to PCCW – an ISP in Hong Kong – which then propagated the false routing information across the whole Internet. This propagation amounted to a denial-of-service (DoS) attack on YouTube. In a DoS attack, users might not be able to obtain access to the Internet or specific websites. This type of attack is also known as a prefix hijacking attack: the Pakistan Telecom AS “hijacked” all traffic destined to YouTube, which amounted to sending Internet traffic meant for YouTube to Pakistan Telecom instead. This scenario involved two types of indicators: a spike in the number of a number of routes that contain the Pakistan Telecom AS and a spike in the AS advertisements made by Pakistan Telecom. A detailed forensic analysis of the attack was published by RIPE (2008).

After reviewing many indicators in the literature, we identified three that were common to most scenarios and should be observed by any analyst interested in BGP attacks:

1. Number of AS announcements: a sharp increase in the number of announcements is typically a strong indicator of hijacking (irrespective of whether it is malicious or not).

2. AS path length: the length of AS paths (list of systems that a BGP route follows from a given AS to the AS that owns a given prefix). During attacks AS path length increases. An analyst can observe the baseline behaviour to determine a typical AS path length and then use it to set a threshold, above which an alert should be thrown.

3. Multiple-origin AS (MOAS) conflict: more than one AS is claiming to be the owner of a given prefix. Any prefix should only be owned by one AS.

*Operationalization of indicators and alerts (automation)*
Figure 1 is a high-level flow diagram showing the main modules of the analysis platform that was constructed and the stages of information flow through the different modules. There are three main stages:

1. Input: collection of real-time data. In the BGP case study, BGP traffic is obtained from public data sources known as RIPE collectors, which archive BGP traffic data from around the world. The platform can either process BGP traffic obtained from collectors directly or use data replayed from an existing case file. The latter is important for training and validation purposes, as well as for forensic analysis of a particular attack.

2. Processing: extract, process, and dispatch features (i.e., key characteristics) of the data in real time (e.g., BGP announcements with information about the time of the announcement, the origin AS, and the AS path). The extracted features are sent to a message broker (Kafka), which will dispatch the information to different internal modules. MongoDB stores the features in a database, which will be used during

# Combining Exploratory Analysis and Automated Analysis for Anomaly Detection

*Ahmed Shah, Ibrahim Abualhaol, Mahmoud Gad, and Michael Weiss*

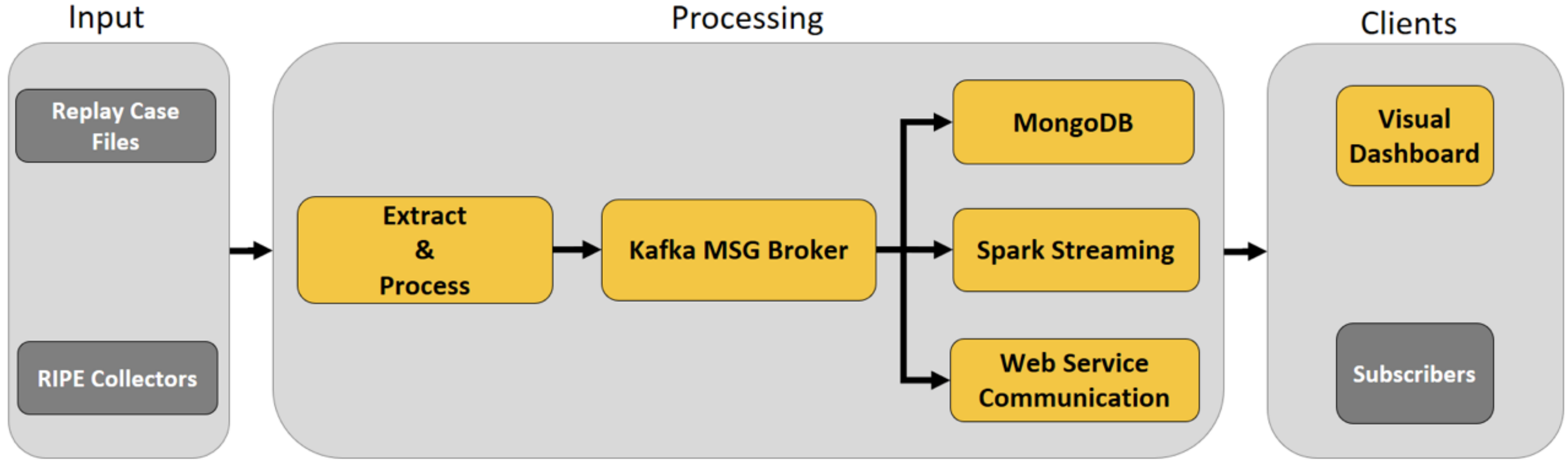

**Figure 1.** High-level architecture of the analysis platform

visualization. Apache Spark conducts further processing such as computing running averages of indicators or comparing indicators to thresholds. Web service communication provides an interface to external analytics systems.

3. Clients: clients include a visual dashboard – where alerts and indicators are visually presented to the user – or external systems that can subscribe to alerts.

*Visualization (exploration)*
Figure 2 shows the user interface (visual dashboard) of the analysis platform with the various visualization components. The visual dashboard contains three main sections:

A. Live Monitor: provides a simple status summary of real-time data stream ingestion.

B. Configure: a command-line-based interface for setting configuration parameters for controlling the input data stream (e.g., which IP prefix to monitor) and setting indicator thresholds.

C. Drill Down: provides a visual interactive dashboard on the data being ingested. This includes displaying recent alerts and providing interactive visualizations of the context of a given alert using timelines, histograms, and other graph types of the indicators that are being monitored.

Through the drill-down capability, the analyst can explore the context of a particular alert. They can zoom into a particular time range, showing only events and data related to that time interval, such as around a spike in a given indicator (e.g., the number of AS announcement). They can see when a given indicator is either unusually high or low by selecting the corresponding value or value range in a histogram component, upon which the other visualization components will be updated to show only corresponding values. For example, selecting just the high values for AS path length will reveal which AS and which prefixes were associated with long AS path lengths. Given that AS path lengths are generally short, a long AS path length may indicate a hijacking attack. By inspecting the origin AS of a long AS path, the analyst can quickly conclude which AS might be the source of the attack.

Figure 2 shows the results of the analysis platform replaying the YouTube 2008 IP Prefix hijacking case. The "number of updates" graph shows that there is a long period of time, from approximately 12:00am to 6:00pm, when updates are infrequent. For an analyst, this stable network activity could be considered a baseline that indicates that nothing beyond normal activity is occurring. When the IP prefix hijacking occurred (at approximately 6:30pm), there was a large increase in the frequency of updates, which may indicate an anomaly that the analyst should explore.

## Trade-Offs between Exploratory Analysis and Automated Analysis

Figure 3 illustrates the interplay of automation and visualization. Automation (on the left) is about creating rules according to which real-time alerts will be raised. Alerts will be shown to an analyst in a dashboard. Visualization (on the right) is about providing the analyst with the ability to interactively explore the data associated with alerts (e.g., focus the analysis on specific time ranges or examine at which times a given indicator displayed unusually low or high values). The exploration of data might suggest patterns in the data (e.g., spikes

# Combining Exploratory Analysis and Automated Analysis for Anomaly Detection

*Ahmed Shah, Ibrahim Abualhaol, Mahmoud Gad, and Michael Weiss*

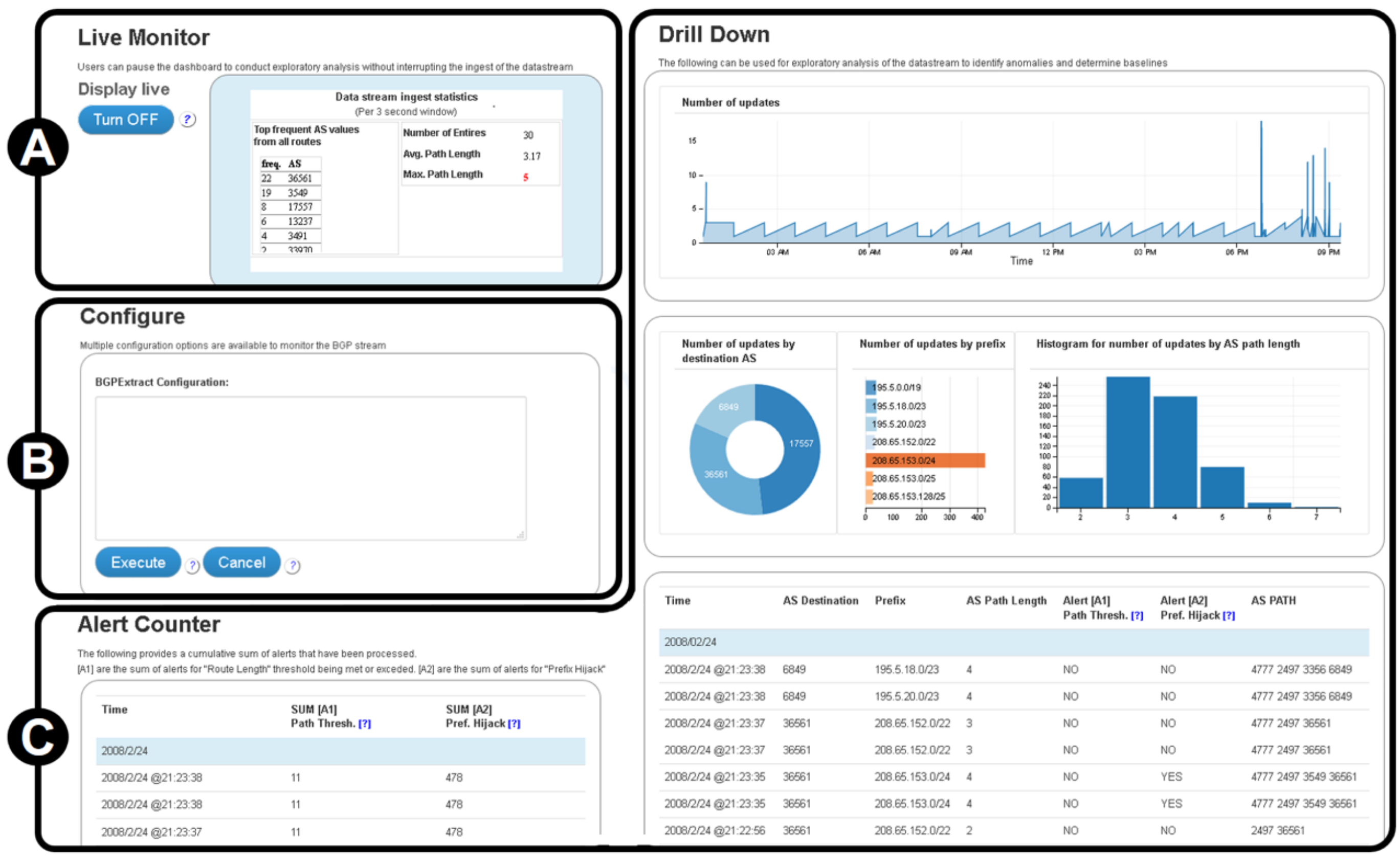

**Figure 2.** User interface of the prototype analysis platform

in a given indicator) that may indicate potential attacks and should be codified into rules (e.g., new or revised thresholds attached to an indicator). This exploration helps build confidence in both the correctness and meaningfulness of the alerts.

On the one hand, we want the alerts and indicator thresholds to be correct. Only then analysts can be expected to rely on them. For example, when showing the AS involved in a prefix highjack attack, the designer of the dashboard may inadvertently be showing destination AS, rather than origin AS. In the case of a prefix highjack, however, only the origin AS will provide insights into which AS may be the source of the problem (such as the Pakistan Telecom AS in the YouTube scenario described in Box 1 and shown in Figure 2). A careful comparison of a known scenario against the values of the indicators in the dashboard can help detect such design errors.

On the other hand, we want the information provided to analysts to be meaningful. For example, if the threshold for an alert is set too low, too many alerts will be generated, overwhelming the analysts. Again, it may be difficult to determine the right threshold beforehand. However, by exploring the data, the analysts will be able to identify typical value ranges and thus suggest appropriate thresholds.

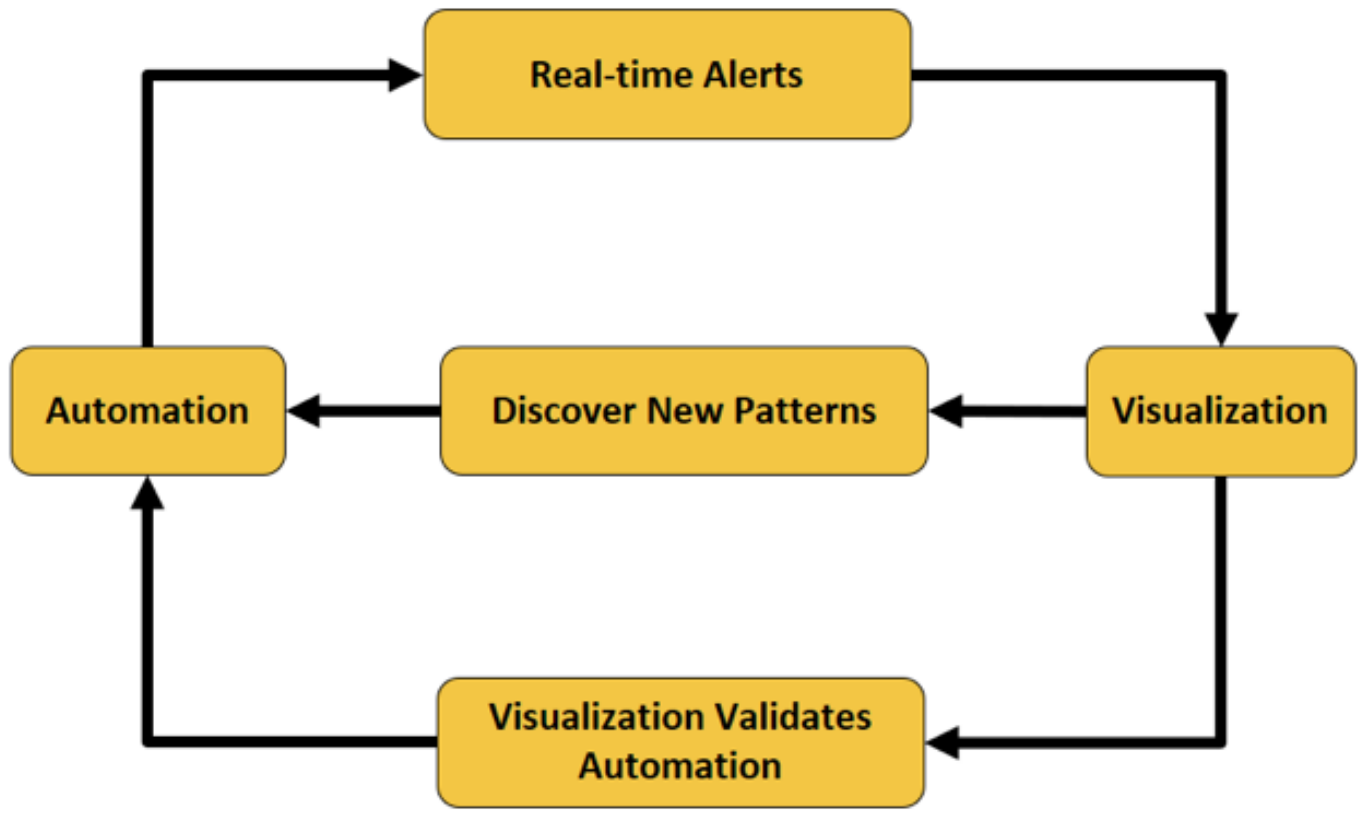

**Figure 3.** Interplay of automation and visualization

# Combining Exploratory Analysis and Automated Analysis for Anomaly Detection

*Ahmed Shah, Ibrahim Abualhaol, Mahmoud Gad, and Michael Weiss*

During initial development of the analysis platform there was a focus on developing automation rules. When visualization components were added, relationships between indicators and anomalous behaviour in case studies became easier to identify, which then set the path for developing relevant automation rules. Automated approaches for anomaly detection should, therefore, be combined with preliminary explorations of the observed environment and data.

## Conclusion

In this article, we explored the interplay between exploratory analysis and automated analysis. We described an experimental system for monitoring real-time data streams that combined exploratory analysis and automated analysis. The prototype incorporated both traditional rule-based mechanisms for detecting anomalies in data streams and interactive tools for discovering new anomalies and validating detection rules. Developers of real-time security monitoring systems can take the lessons from this research to reinforce the importance of how exploration and automation complement each other. Future work may include creating a real-time security information management system (SIEM) that uses machine learning to identify baseline patterns and potential attack patterns for processing data streams while also developing visualization components to tune algorithm accuracy.

## About the Authors

**Ahmed Shah** holds a BEng in Software Engineering from Lakehead University in Thunder Bay, Canada, and a MEng in Technology Innovation Management from Carleton University in Ottawa, Canada. Ahmed has experience working in a wide variety of research roles at the VENUS Cybersecurity Corporation, the Global Cybersecurity Resource, and Carleton University.

**Ibrahim Abualhaol** is a Research Scientist at Larus Technologies and an Adjunct Professor at Carleton University in Ottawa, Canada. He holds a BSc, an MSc, and a PhD in Electrical and Computer Engineering. He is a senior member of IEEE and a Professional Engineer (P.Eng) in Ontario, Canada. His research interests include real-time big-data analytics and its application in cybersecurity and wireless communication systems.

**Mahmoud M. Gad** is a Research Scientist at the VENUS Cybersecurity Corporation. He holds a PhD in Electrical and Computer Engineering from the University of Ottawa in Canada. Additionally, he holds an MSc in ECE from the University of Maryland in College Park, United States. His research interests include big-data analytics for cybersecurity, cyber-physical system risk assessment, cybercrime markets, and analysis of large-scale networks.

**Michael Weiss** holds a faculty appointment in the Department of Systems and Computer Engineering at Carleton University in Ottawa, Canada, and he is a member of the Technology Innovation Management program. His research interests include open source, ecosystems, mashups, patterns, and social network analysis. Michael has published on the evolution of open source business, mashups, platforms, and technology entrepreneurship.

# Combining Exploratory Analysis and Automated Analysis for Anomaly Detection

*Ahmed Shah, Ibrahim Abualhaol, Mahmoud Gad, and Michael Weiss*

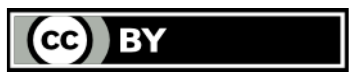